\documentclass[preprint,floats,eqsecnum,aps,tightenlines,showpacs]{revtex4}
\usepackage{epsfig}

\newcommand{\be}{\begin{equation}}
\newcommand{\ee}{\end{equation}}
\newcommand{\bea}{\begin{eqnarray}}
\newcommand{\eea}{\end{eqnarray}}

\newcommand{\nn}{\nonumber}

\begin{document}


\title{Solution of Dynamical  Chiral Symmetry
Breaking in  Minkowski Space\\
{\it  Linearized approximation}}

\author{Vladim\'{\i}r \v{S}auli}

\affiliation{Department of Theoretical Physics,
Nuclear Physics Institute, \v{R}e\v{z} near Prague, CZ-25068,
Czech Republic}

\begin{abstract}
Chiral symmetry breaking and mass generation is studied in a vectorial, confining
asymptotic free gauge theory. Using the Schwinger-Dyson equation  in  improved ladder 
approximation, we calculate the fermion propagator in the whole Minkowski space. 
The estimate for $f_{\pi}$ and the dependence of physical mass on a 
 coupling strength is provided. We focus on the extraction of spectral function of fermion propagator in the strongly coupled regime. Our calculations indicate that up to the crossover between walking and QCD-like running dynamics, the real pole of the propagator is not excluded
and very likely it is actually developed at zero temperature theory.

\end{abstract}

\maketitle
%

\section{Introduction}

Chiral symmetry plays  a crucial role in  particle physics.
Our recent understanding of low energy $E\simeq \Lambda_{QCD}$ hadron physics is heavily based on this phenomenon. The observed pions are naturally identified with the approximatively
 massless Nambu-Goldstone modes of broken $SU(2)_L\times SU(2)_R$ chiral symmetry. The associated
quark mass generation is driven by the QCD strong interaction of  light quarks via the colored gauge boson of  $SU(3)_c$ Yang-Mills  theory. Although  experimentally observable small current masses enter the QCD Lagrangian, the huge amount of constituent quark mass  is generated dynamically.

There are also interesting proposals to solve hierarchy problem of
the Standard Model without point-like Higgs, 
which invoke strong interactions. They include
Technicolor models \cite{FASUS1981}, Topcolor assisted Technicolor models
\cite{TOPCOLOR}, Extended Technicolor Models \cite{DISU1979,EILA1980} with walking effective coupling \cite{HOLDOM1985}
  and even the extradimensional one
\cite{GHTY2002}.  In  realistic versions of  models,  the  part of the chiral group is a subgroup of
standard model $SU(2)_w\times U(1)_Y$, while the breaking of chiral symmetry automatically implies
spontaneous electroweak  symmetry breaking.  Hence these models are very ambitious,
as they should explain the observed masses of fermions, but they must give
the correct masses to $W^{\pm},Z$ as well. For a rather recent review of electroweak symmetry breaking in Technicolors see \cite{REWIEV}.

In the above mentioned models  the particles masses are mostly (if not fully)
generated via loops. If  the coupling constant is small enough we can say that
the chiral symmetry plays the role of 'custodial symmetry' which prevents the 
particle from receiving a mass. Hence relatively strong coupling is needed,  the dynamical fermion mass generation  is therefore intrinsically non-perturbative phenomenon of  quantum field theory. There are some challenging problems connected to the  singularity
structure of Green's functions in strong coupling models like QCD or
Technicolors: e.g., questions about the presence and the position of  propagator poles.
In view of recent achievements in lattice calculations and in numerical
results obtained by employing Schwinger-Dyson equations  in the
Euclidean space,  solutions obtained in the the whole Minkowski space
should bring new insights for these models.

Let us mention some background and related work. The gap equations have been used for many years to study spontaneous chiral symmetry breaking in field theories \cite{SCALING,GAUGE,KLUCI,CHIRALY,HAKUYA2005} 
Solving the Schwinger-Dyson equation  directly in Minkowski space has
an advantage of providing the solution also for timelike momenta,
which are not directly accessible in Euclidean formalism. Up to date, only few attempts or suggestions \cite{BICUDO,ADBISA,ARRBRO2003,BLUJOH1998}  to solve the problem of dynamical  chiral symmetry breaking in the whole Minkowski space are known. 
In our presented approach, we assume and employ dispersion relation for the dynamical mass function $\Sigma$.
However we should note that the resulting integral representation for the fermion propagator cannot be identified with the standard Lehmann spectral representation.
Thus giving up  the assumption of a Lehmann spectral representation for propagator, which seems to be unavoidable due to the high momentum  asymptotic of the fermion correlators, we  obtain perfectly stable solutions even far away from the chiral phase transition. We  compare our Minkowski solutions with ones obtained
independently in Euclidean space.

The paper is organized as follows. In the next section    we will introduce the model and formalism, in  section \ref{tri}   we explain the solution of gap equation in Minkowski space, in  section \ref{ctyri} we will  check the linear approximation employed. In   section \ref{pet}  physical scaling law will be provided, i.e.,  the dependence  of the pole mass on the coupling. We   discuss further possible progress in the Conclusion and Outlook. 

\section{ \label{dva} Toy model and basic formalism}

As our toy model consists of just one interacting quark,  there is just  one massless
pion associated with broken chiral symmetry $U(1)\times U(1)/U(1)$. Once the chiral symmetry is spontaneously broken, the massless pole appears in the fermion-boson vertices
and the associated current of broken generator induced non-zero transition between vacuum
and the pion itself. In that case the pion decay constant
\be
<0|j_5^{\sigma}(0)|\pi(k)>=if_{\pi}k^{\sigma}
\ee 
can be estimated by  the Pagels-Stokar formula \cite{PASTO}
\be \label{pasto}
f_{\pi}^2=-4iN_c\int\frac{d^4p}{(2\pi)^4}\frac{\Sigma(p^2)}{(p^2-\Sigma^2(p^2))^2}
\left[\Sigma(p^2)-\frac{p^2}{2}\frac{d\Sigma(p^2)}{dp^2}\right] \, ,
\ee
where $\Sigma$ is  mass function  defined through the fermion propagator:
\begin{equation}
S(p)=\frac{Z}{\not p-\Sigma(p^2)} \, .
\end{equation}
wherein  we take renormalization wave function $Z(p)=1$ from now
since its momentum dependence is unimportant in Landau gauge which we will use
in our calculation.

The basic input into the relation for the pion decay constant  is the dynamical mass $\Sigma$,
which we shall  assume here to satisfy the dispersion relation:
\be 
\Sigma(p^2)=\int\limits_{0}^{\infty} d\omega\,
\frac{\rho(\omega)}{p^2-\omega+i\epsilon} \, .
\label{dr}
\ee

In that case we approximate the $\Sigma^2$ in the denominator of $S$
by the pole value $m^2$, i.e. 
\be \label{aprox}
S=\frac{\not p+\Sigma(p^2)}{p^2-\Sigma^2(p^2)}\rightarrow
\frac{\not p+\int\limits_{0}^{\infty} d\omega\,
\frac{\rho(\omega)}{q^2-\omega+i\epsilon} }{p^2-m^2+i\epsilon}
\ee
where the pole mass is determined self-consistently as
\be \label{pole}
\Sigma(m^2)=m  \, .
\ee
As we will see in the next section the  determination of  the function $\rho$ and  physical mass $m$
is not difficult task in our denominator approximation.

The dynamical mass is approximately constant function in the infrared $p^2<\Lambda$, while it decreases like power of momentum 
\be \label{padam}
\Sigma\simeq \frac{1}{p^2}
\ee
for $p^2>>\Lambda^2$ in an asymptotic free theory characterized by the logarithmically behaving running coupling above the scale $\Lambda$.
Recall the QCD scaling here: $E\simeq \Lambda_{QCD}(=400 MeV)\simeq \Sigma(0) (=350 MeV) \simeq4f_{\pi}$. 
In our model the selfenergy  behaviour (\ref{padam}) is achieved a bit  faster  as a consequence of the used  Pauli-Villas regulator.  When $\Sigma$ approaches $\Lambda$, then $\Lambda$ represents the confinement scale \cite{WITTEN} and confinement goes hand by hand with the spontaneous chiral symmetry breaking.  Unfortunately, the denominator approximation method of this paper are then no longer useful in this case. Nevertheless, the proposed denominator approximation  should be rather good \cite{CHIRALY} for walking (slowly running) theories when  $m^2<<\Lambda^2$.
 
Here, a short but important digression is in order. 
Clearly, the propagator given by formula (\ref{aprox}) exhibits a real-value pole.
This gives rise the natural question of using a Lehmann representation (see the first two formulas in the Appendix) for this purpose. This has been attempted in the paper \cite{ADBISA} wherein it was found that a spectral representation with a single real pole part explicitly divided is applicable only with crude restrictions. Indeed, it gives reliable result only up to the scale $\Lambda$ and for the couplings which are very close to the critical value $\alpha_c$. It seems that $1/p^2$ behaviour of the $Tr S(p)$ (Tr is taken over the Dirac indices) is  dictated  too strongly by Lehmann representation, which contradicts  $1/p^4$ of $Tr S(p)$ decrement typically found in  the Euclidean space studies. It was also shown in  \cite{ADBISA} that the propagator looses its assumed analytical property in this case. It is  stated  without proof here, that the ultraviolet asymptotic of the correlators strongly suggest the absence of Lehmann representation for propagators in  models with the dynamical  chiral symmetry breaking. Perhaps and with  great care, the Lehmann representation can be used as a low energy approximation.

In addition we neglect the terms with the derivatives and substitute our
integral Ansatz (\ref{aprox}) into the Pagels-Stokar formula (\ref{pasto}) for the pion decay constant. Doing this explicitly we get
\be \label{ee}
f_{\pi}^2=-4iN_c\int_0^{\infty} dx_1 dx_2 \int
\frac{d^4p}{(2\pi)^4}\frac{\rho(x_1)\rho(x_2)}{(p^2-m^2)^2(p^2-x_1^2)(p^2-x_2^2)}.
\ee
After the momentum integration we arrive into the following formula:
\be \label{eee}
f_{\pi}^2=-\frac{N_c}{4\pi^2}\int_0^{\infty} dx_1  dx_2 
\frac{\rho(x_1)\rho(x_2)}{x_1-x_2}
\left[\frac{\ln{(m^2/x_1)}}{x_1^2-m^2}-\frac{\ln{(m^2/x_2)}}{x_2^2-m^2}\right]
\ee
This can be further approximated since one can easily recognize that up to the presence of log's the double integral factorizes and turns to be square of the pole masses $m^2$ exactly. For this estimate  we  approximate slowly varying log's in (\ref{eee})by a constant $K=\ln{(m^2/x_{max})}$ where $x_{max}$ is the position of  $|\rho|$ function global maximum. Anticipating the typical behaviour of  $|\rho|$ it 'explodes' from the threshold where it is zero to its maximum which lies not  so far. Doing this explicitly and using
(\ref{pole}) we immediately  get:
\be \label{eeee}
\frac{f_{\pi}}{m}=\frac{\sqrt{K N_c}}{2/\pi} \, .
\ee
Numerically $K\simeq 0.5$ when $\Sigma(0)\simeq \Lambda$ which by about 50 percentage underestimates an  experimentally well tested QCD result. It is not so bad estimate because of denominator approximation, Eq. (\ref{eee}) represents at least an interesting formula which relates the physical pole mass $m$, the absorptive part of dynamical mass function  $\rho$ and the decay constant of the Nambu-Goldstone boson. Finally notice that it is impossible to obtain such result within the use of Lehmann representation since the formula appears to be ultraviolet divergent in that case. 

Of course, in any case $f_{\pi}$ can be calculated purely from the Wick rotated results on $\Sigma$ (see Appendix ) and this is not  clearly a
manifest quantity where the knowledge of the timelike structure of the propagator is necessary.
Various timelike form factors  are the right possible  cases. To be able to calculate them from
 first principles in QCD the nonperturbative  knowledge of $\Sigma$ 
 in the whole Minkowski  is required.

 The vectorial interaction 
with gluons leads to the gap equation which in the Landau gauge and one skeleton loop approximation reads
\bea \label{zdar}
\Sigma(p^2)&=&-ig^2C_2(R)\int\frac{d^4q}{(2\pi)^4}\gamma_{\alpha}
\left[-g^{\alpha\beta}+\frac{(p-q)_{\alpha}(p-q)_{\beta}}{(p-q)^2}\right]\,
G(p-q)S(q)\gamma_{\beta}\, . 
\eea
where $C_2(R)=(N^2-1)/2N$ for fermions in fundamental representation of SU(N).
To implement the running coupling, a  very simple Ansatz for the product of gluon and  vertex form factors has been used in \cite{ADBISA}. It was chosen such that for a large Euclidean momenta
such running coupling $\alpha_{run}$ behaves like 
\be
\alpha_{run}(q^2)\simeq \frac{\alpha_{run}(0)}{1+\frac{q^2}{\Lambda^2}}
\ee
which is continuous modification of sharp cutoff used in a recent studies \cite{KURACHI2006}.

Implementation of  $\alpha_{run}$ is  easily achieved by adding negative term to the usual
boson propagator,i.e.
\be
G(l)=\frac{1}{q^2-m_g^2+i\epsilon}-\frac{1}{q^2-{\Lambda}^2+i\epsilon}
\ee
in Eq. (\ref{zdar})
where we have introduced also an effective gauge boson mass $m_g$.

Further note, that we prefer to have only one dominant scale and we do not attempt to solve the model in  all its possible peculiar phases. For this purpose we fix
the ratio such that 
\be
\frac{\Lambda^2}{m_g^2}=10 \, .
\ee
Throughout this paper we will express the dimensionfull quantities in the units of 
$m_g=1$ and/or $\Lambda$ if stated explicitly. 

Let us stressed at the end of this section that  the scale $\Lambda$ mimic $\Lambda_QCD$
or characteristic scale of  an asymptotic free gauge theory and it has nothing to do 
with the renormalization/regularization procedure which does not take place here.  Asymptotic freedom is necessary condition which make our calculation meaningful. In the real QCD
the high energy modes are damped by logarithm which is the main difference here. As we will see,
the running (or walking) of the effective coupling largely affects the asymptotic behaviour of the fermion mass function, it behaves like $1/p^4$ while 
the mass of chiral quarks behaves like $1/p^2$ for $p^2>>\Lambda^2_{QCD}$

\section{\label{tri} Solving gap equation for {\large $\rho$} }

The simple integral representation Ansatze allows us to convert the complicated equations with singular kernel to the real regular equation for the function $\rho$.
The key point is the analytical momentum integration in the gap equation
  with self-consistent arrival to the desired dispersion relation, i.e.,   into the  form of propagator   which  we have assumed and from which we have started the calculation. 
The only assumption we need is the  uniqueness of the dispersion for $\rho$.

Applying  the trivial algebraic identity on integral representation of $S$   
\bea
\int d x \frac{\rho(x)}{(q^2-x+i\epsilon)(q^2-m^2+i\epsilon)}
=\int d x \frac{\rho(x)}{(m^2-x)}
\left[\frac{1}{q^2-m^2+i\epsilon}-\frac{1}{q^2-x+i\epsilon}\right]
\eea
and abstracting from various prefactors,  the integral kernel reduces to the differences
of one loop perturbation theory integrals, each of them being regularized by Pauli-Villars. 
Consequently we 
can immediately use this result. 

The momentum integration is very straightforward since it follows standard tricks.  E.g., the Feynman parameterization is a convenient way to perform it explicitly. At the end the Feynman parameter is converted into the integration variable of an appropriate dispersion relation. The integral is finite and does not require any subtractions as it must  in the theory without explicit mass term. 
For convenience, the relevant formula can be found for instance in the paper
\cite{LACO} or in any standard textbook.

Hence after  substituting the integral Ansatz (\ref{aprox}) into the gap equation, making a trace and  integrating over the momentum it  leads to the following  expression for the dynamical mass:
\begin{eqnarray}
\label{pispunta}
\Sigma(p^2)&=& \int d\omega \frac{\rho(\omega)}{p^2-\omega+i\epsilon}
\, , \nn \\
\rho(\omega)&=&\frac{\alpha^*}{(4\pi)}\int d x \frac{\rho(x)}{m^2-x}
\left[X_0(\omega;m_g^2,m^2)- X_0(\omega;{\Lambda}^2,m^2)\right.
\nn \\
&-&\left.X_0(\omega;m_g^2,x)+ X_0(\omega;{\Lambda}^2,x)\right]\, ,
\label{ie} \\
X_0(\omega;a,b)&=&\frac{\sqrt{(\omega-a-b)^2-4ab)}}{\omega}
\Theta\left(\omega-(\sqrt{a}-\sqrt{b})^2\right)\, ,
\end{eqnarray}
where $\Theta$ represents for the  usual Heaviside step function and we use following shorthand notation:
\be
\alpha^*=\frac{C_2(R)g^2}{4\pi}.
\ee

Assuming of uniqueness  DR  the Eq. (\ref{ie}) represents a homogeneous integral equation to be solved. The Eq. (\ref{pispunta}) has a regular kernel. In general the homogeneous equation needs a suitable 'normalization condition'. Here this can be easily identified with  the 
 condition of on-shellness (\ref{pole}), i.e.
\be  \label{identit}
\int_{(m_g+m)^2}^\infty d x \frac{\rho(x)}{m^2-x}=m
\ee
where we have explicitly indicated the branch point as  dictated by the arguments of step functions in (\ref{ie}). Using Rel. (\ref{identit}),  the equation to be actually solved numerically now reads:
\bea
\rho(\omega)&=&\frac{\alpha^*}{(4\pi)}
\left[m\left(X_0(\omega;m_g^2,m^2)- X_0(\omega;{\Lambda}^2,m^2)\right)\right.
\nn \\
&-&\int d x\frac{\rho(x)}{m^2-x}  
 \left.\left(X_0(\omega;m_g^2,x)- X_0(\omega;{\Lambda}^2,x)\right)\right]\, .
\label{ie2} 
\end{eqnarray}

Using  the method of iterations it is found that the integral equations is perfectly convergent and stable for almost any combinations of 
$\alpha^*$ and $m$ (without imposed (\ref{identit})).  Hence the actual search of the true solution is very straightforward. For a given $\alpha^*$ we scan suitably chosen interval of $m$  in such a way that the identity (\ref{identit}) is achieved with required accuracy.

For the purpose of  comparison we solve Euclidean gap equation.
After the Wick rotation and the angle integration it can be cast into the 
following form:
\begin{eqnarray}
\label{euclid} M(x)=m_0+\frac{\alpha^*}{8\pi x}\,
\int_0^{\infty}d y \, M(y)
\frac{-m_g^2+\Lambda^2+\sqrt{\lambda(x,y,m_g^2)}-
\sqrt{\lambda(x,y,{\Lambda}^2)}}{y+M^2(y)} \, ,
\end{eqnarray}
where the positive variables $x=p_E^2=-p^2,y=q^2_E=-q^2$ 
and the dynamical mass is simply $M(x)=\Sigma(p^2)$.
$\lambda$ represents standard Kallen triangle function
\be
\lambda(x,y,z)=x^2+y^2+z^2-2xy-2xz-2yz
\ee

\begin{figure}
\centerline{\epsfig{figure=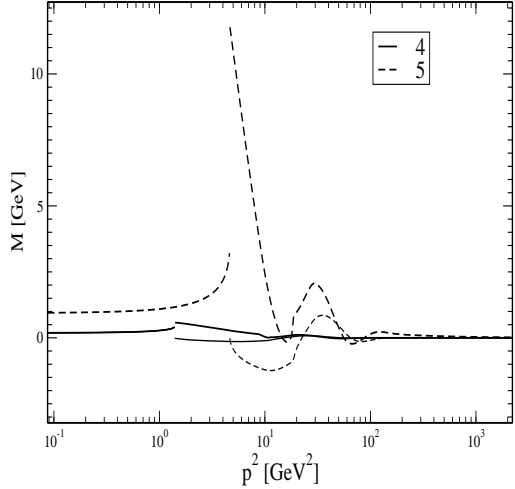,width=8truecm,height=8truecm,angle=270}}
\caption[caption]{Dynamically generated mass functions at timelike
regime. Dashed(solid) line represents the solution for $\alpha^*=5$($\alpha^*=4$).
Thick (thin) line stands for dispersive (absorptive -$\rho$-) part respectively. } 
\label{timelike}
\end{figure}

\begin{figure}
\centerline{\epsfig{figure=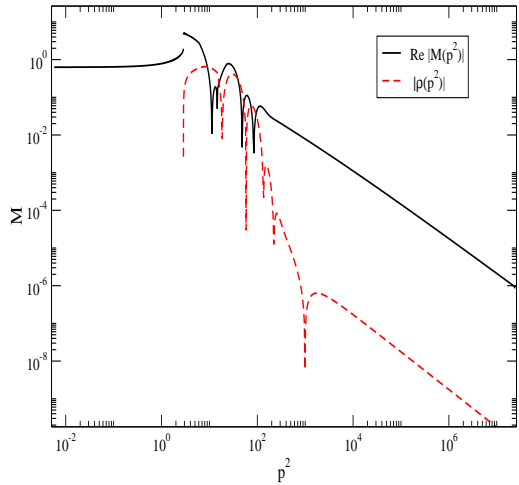,width=8truecm,height=8truecm,angle=270}}
\caption[caption]{Log-log display of the absolute values of real an imaginary parts of dynamical mass for $\alpha^*=5$.} \label{spektrum}
\end{figure}

First, let us discuss the  timelike momentum behaviour.
The resulting absorptive part of the dynamical mass function starts with negative infinite derivative from a zero at the threshold $\omega^{1/2}=m_g+m $ reaching the first most robust peak. 
Then after going through two additional local maxima (minima) it behaves exactly like
\be
\rho(p^2)\simeq \frac{1}{p^2}
\ee
for $ p^2>>\Lambda^2$ without any additional oscillation.
The resulting mass function is displayed in 
Fig. \ref{timelike} for two values of the coupling $\alpha^*=4$ and $\alpha^*=5$. Most interestingly, to see a detailed  behaviour
we plot the  absolute value of  real (dispersive) and absorptive (imaginary) part of the dynamical mass in Fig. \ref{spektrum}.
 We argue that the observed "resonant like behaviour"  Green's
functions is an inherent property of strong coupling -supercritical- dynamics.
That this must be so for higher point Green's functions due to the formation of hadronic bound states and resonances is an experimental fact and indeed an almost trivial statement in QCD.
Recall that the similar thing could happens to propagators of quarks is non-trivial observation. 
Recall that, this is obtained even without nontrivial inputs from higher order corrections involving mainly the dressed vertices which are  connected directly with the physical spectra e.g. the lowest lying meson states $\pi, \rho, \omega, \sigma ..$.   
 We should stress at this place that there are only tiny numerical errors ( we use for integration  number of 990 Gaussian mesh points  in the both cases presented, however using  few hundreds with the suitable  density is always justified). We mention here also that the principal value integral over the $\rho$ has been calculated analytically, the result of which is presented in the Appendix for completeness. The cancellation among negative and positive  contributions in the integral  (\ref{dr}) gives the correct -$p^{-2}$ - ultraviolet decreasing  
log moderated behaviour of the dispersive part of the dynamical mass function.

\begin{figure}
\centerline{\epsfig{figure=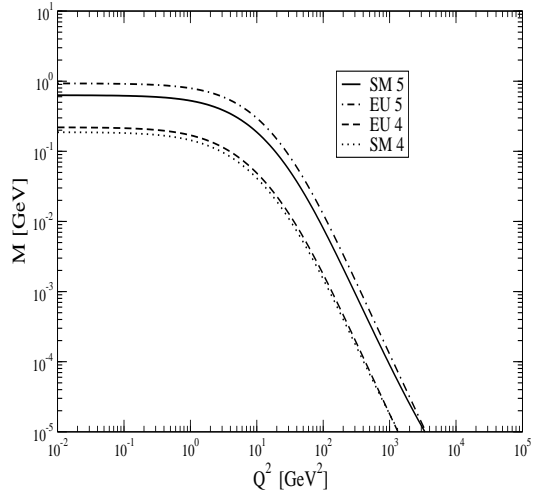,width=8truecm,height=8truecm,angle=270}}
\caption[caption]{Dynamical mass function in the spacelike regima of momentum.
Spectral Minkowski solutions are labeled by SM , the solutions obtained in the Euclidean space are labeled by EU.}
 \label{hmota}
\end{figure}
The dynamical mass function for the spacelike momenta is displayed in Fig. \ref{hmota}.
Being below the branch point the integration washout the oscillating behaviour and the dynamical mass function is a smooth function. We can see in Fig. \ref{hmota}
that not so far from the critical coupling $\alpha_c^*=3.8$  Minkowski and Euclidean results reasonably agree with each other. The numerical results of our studies point  towards an analytical structure of the 'quark' propagator with a real pole and branch point singularity on the real axis.   Close to the chiral phase transition, the suggested nature of this singularity is determined with good confidence. 

The impact of difference between walking and running $SU(N)$ gauge theories has been recently investigated in Refs. \cite{KURACHI2006}. 
Our results give additional information:  vanishing difference between Euclidean and Minkowski solution suggests that there are no additional complex singularities in the complex plane of the momenta. 
In  Technicolor slang: in walking theory the fermion propagator singularity is dominated by a real pole and associated particle mode can be unconfined.
Disappearance of the real propagator pole would be a certain indication of confinement
in any  model which is heavily based on the walking behaviour of the coupling constant.
 Recall the possibility of chiral symmetry breakdown and the absence of confinement is  supported by an  arguments in \cite{CHIRALY}.

\begin{figure}
\centerline{\epsfig{figure=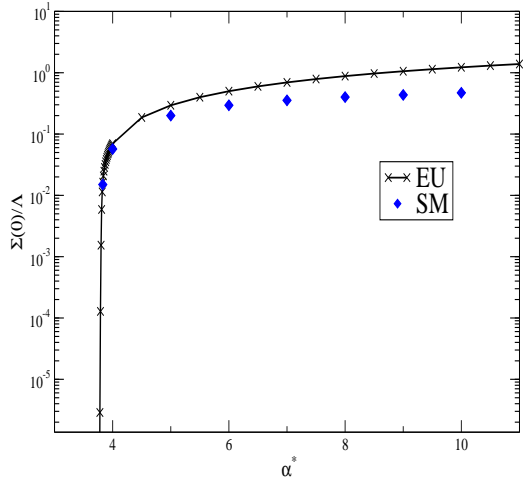,width=8truecm,height=8truecm,angle=270}}
\caption[caption]{Scaling laws as it is seen in the Euclidean and Minkowski approximations.} \label{scaling}
\end{figure}

In Fig. \ref{scaling} we show the solution for dynamical mass as a function of $\alpha^*$.
In the  walking regime $(1-1.1)\alpha_c^*$ it can be parametrized by the known  
\cite{SCALING,CHIRALY} scaling low
 \be
\Sigma(0)\simeq C\cdot \Lambda e^{-\left[k(\alpha^*/\alpha_c^*-1)^{-1/2}\right]} \, ,
\ee
where $C=O(1)$and $ k\simeq \pi$ .

Increasing the coupling and leaving thus the walking coupling domain   and slowly approaching
QCD limit $\Sigma\simeq \Lambda$  a certain discrepancy appears.
In present, the nature of the all amount of the
discrepancy cannot be determined with confidence.
However we point out the main part is  due to the linearized approximation (\ref{aprox}). 
The other, more speculative reason, could be that the pole of the
propagator, located for subcritical case on the real axis, moves
gradually into the complex plan receiving non-negligible
imaginary part. Hence the  integral representation employed for
our calculations would become inappropriate.

\section{\label{ctyri} Checking the approximation (\ref{aprox})}

At present we do not know any reason
why the spacelike Euclidean calculation should differ from the 'true' 
Minkowski space calculation on the real negative axis of four-momentum. Hence the calculation performed in the Euclidean formalism can serve as a good guidance.
To observe how much the neglect of a running mass in the denominator (\ref{aprox}) affects the behaviour in spacelike domain, we make a similar approximation in the Euclidean space counterpart.
For this purpose we replace in Eq. (\ref{euclid}) the appropriate part of the propagator  follows
\be
\frac{M(y)}{y+M^2(y)}\rightarrow \frac{M(y)}{y+M^2(0)} \,
\ee
hoping that $M(0)$  lies sufficiently close to the true pole value $m$. 
Note, albeit we have estimated  $m$ in our Minkowski treatment 
we cannot use this value, since this is already underestimated
just due to the discussed approximation. 

\begin{figure}
\centerline{\epsfig{figure=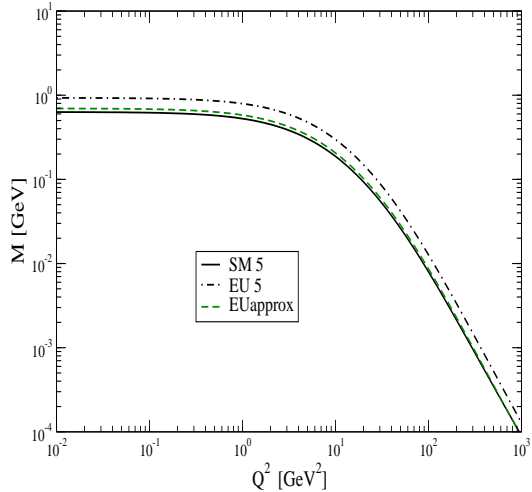,width=8truecm,height=8truecm,angle=270}}
\caption[caption]{Checking denominator approximation as described in the text.
 The coupling strength is $\alpha^*=5$ and the obtained pole mass is $m=0.695m_g$}
 \label{hmota3}
\end{figure}

The solution  is shown in Fig. \ref{hmota3} for  $\alpha*=5$.

The corresponding approximation is labeled by the dashed line and much better agreement with the Minkowski approximation is achieved. It leaves  little space for other effects and indirectly indicates that the position of propagator singularities actually lies on the real axis.
We should mention that the approximation (\ref{aprox}) exhibits a similar effect for all studied value
of $\alpha^*$. On the other side, the splitting of the obtained scaling law functions becomes rather large  when $m/\Lambda\simeq 1$. To avoid some speculations we conclude that it simply calls to go beyond the 
approximation employed.

\section{\label{pet} Physical scaling law}

\begin{figure}
\centerline{\epsfig{figure=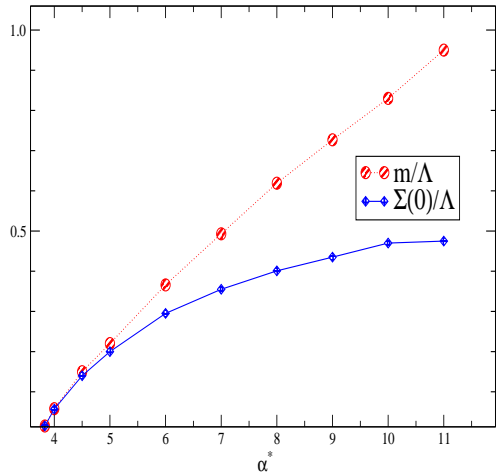,width=8truecm,height=8truecm,angle=270}}
\caption[caption]{Comparison of physical scaling law with $\Sigma(0)$ in Minkowski  approximations.} \label{scaling2}
\end{figure}

The pole mass of the propagator can be  the only   physical observable here. 
In this section we will largely simplify and speculate.
We disregard  the systematic error which follows from the approximations
and assume   the ratio $m/\Sigma(0)$ being reliably estimated.
We leave aside the complicated task of confinement-deconfinement-chiral phases 
\cite{WITTEN,PHASES} which 
lies beyond the scope of this paper and certainly beyond the ability of the
toy model used here. We will simply assume that the modeled propagator is a part of the theory in which the singularities of the Green's functions persist in the S-matrix (That the opposite can happen is well known). In this case the pole mass of the propagator is the only possible  physical observable here. In a gauge theory, the value of dynamically generated $m$ cannot be affected by a gauge fixing choice and cannot depend on the associated auxiliary parameters. To make
some approximations in the infinite gap equations system is necessary, e.g., the truncation of the equations system,  have always some unwanted impact on the calculation scheme (in)dependence. Many studies have investigated this 
important, but complicated issue. For some discussion in the context of Euclidean Schwinger-Dyson formalism see \cite{GAUGE}. The value  $\Sigma(0)$ is usually believed be an approximately scheme invariant if its value is rather close to the  physical $m$.  Then there follows important message  from the Minkowski study attempted here. 
We present the physical scaling low in Fig. \ref{scaling2} and compare with the more standard infrared value $\Sigma(0)$. 
The observation of large splitting of the lines in   Fig.  \ref{scaling2} is clear. The ratio 
 $m/\Sigma(0)$ deviates from the save  value one rather early.  In fact, approximating $m$ by the 
infrared $\Sigma(0)$ is justified only for slowly running theories.

\section{Conclusion and outlook}

In the ladder approximation of gap equation with 
 the running  coupling modeled  by Pauli-Villars
regulator we obtain  a dynamical mass function in the whole Minkowski space. The momentum space gap equation has been converted to the real regular equation for a real function $\rho$. The procedure is for us surprise perfectly stable for  all couplings studied. We compare the solution with the ones obtained in the Euclidean formalism. The employed approximation works
 rather well for the coupling not so far from the critical one. The Minkowski results appears to be  underestimated  by about 50 percentage  when we get scaling $\Sigma(0)/\Lambda=1$. The physical pole mass has been compared with the infrared one and the physical scaling has been established.

We did not plot the the ratio $f_{\pi}/m$ as a function of $\alpha* (,\Lambda)$.  Using a rough estimate by the  Pagels-Stokar formula we found that up to deep walking limit this ratio has only  weak dependence on the coupling strength $\alpha^*$. It decreases from the value $f_{\pi}/m=0.5$ when $\alpha$ is few percentage above  $\alpha_c^*$ to the value 0.25 when $\alpha\simeq 10$ (QCD limit).

 The absorptive and dispersive parts of the dynamical mass function are not positive definite functions in the timelike regime, while the later decreases monotonically down in the spacelike regime in a shape already known from the Euclidean studies.
The resonant behaviour of Green's function in the timelike regime can be very inherent to the model with strong coupling, especially for the light quarks in real world QCD.
  If such behaviour is  the reality and general feature of the fermion propagator in  chiral breaking phase it calls for a better understanding of its physical impact. The presented toy model hopefully provides a reliable result for slowly running coupling. The application to many questions of electroweak symmetry breaking in Technicolor models is  left to a future work. When one moves far away from walking behaviour, one loses a simplifying  feature of our calculation, namely that the Pauli-Villars form factor in  use does not reliably approximate QCD running coupling. A more elaborated approach suitable for such situations is presently in progress.

\section{Acknowledgments}

I am grateful to Frieder Kleefeld for a carefull reading of the manuscript.

\appendix

\section{Useful formulas}

\begin{center}{\bf Kallen-Lehmann representation:}\end{center}

For fermion propagator in parity conserving theory reads
\begin{eqnarray}
S(p)=\not p\, S_v(p^2)+ 1\, S_s(p^2)=\int_{C} d\omega\,
\frac{\not p\sigma_v(\omega) +\sigma_s(\omega)}{p^2-\omega+i\epsilon}\, ,
\end{eqnarray}
hence assuming real pole, the only possibility is to assume Dirac delta distribution 
as the lowest mode. In that case we can write for instance for $S_s$
\begin{eqnarray} \label{srr}
 S_s(p^2)=\frac{r}{p^2-m^2+i\epsilon}
+\int_{(m+m_g)^2}^\infty d\omega\,
\frac{\sigma_s^c(\omega)}{p^2-\omega+i\epsilon}\, ,
\end{eqnarray}
where $r$ is a residuum, which does not vanish when dynamical chiral symmetry breaking 
is attempted within this formalism.  Recall the Ansatz (\ref{srr}) works rather well when it is 
used in the  Schwinger-Dyson equations' formalism with explicit mass terms and for the subcritical couplings \cite{LACO,SAUADA2003,SAULIJHEP,SAULIRUN}.

\begin{center} {\bf Dispersion relation integral}\end{center}

There are few convenient possibilities how to perform integration in the dispersion relation
for the dynamical mass. Most simple is to use numerical data for $\rho$ and integrate numerically. 
This is the way which we actually follow below the branch point. 
On the other hand, one can  avoid numerical principal value when evaluating $M$ for timelike $p^2>(m+m_g)^2$  by analytical integration. For this purpose we can use the equation (\ref{ie2}) and substitute it into the (\ref{dr}), arriving thus to the following regular integral:
\begin{eqnarray}
Re \Sigma(p^2)&=&\frac{3\alpha}{4\pi} \int d x
\frac{\rho(x)}{m^2-x}
\nn \\
&&\left[
J(p^2,m^2,m_g^2)-J(p^2,m^2,{\Lambda}^2)
-J(p^2,x,m_g^2)+J(p^2,x,\Lambda^2)\right]\, ,
\end{eqnarray}
where we have defined
\begin{eqnarray}
J(p^2,y,z)&=& -\frac{\Theta(-\lambda_p)\sqrt{-\lambda_p}}{p^2}
\left[\frac{\pi}{2}+
arctg\frac{p^2-y-z}{\sqrt{-\lambda_p}}\right]+\frac{1}{2}\ln(16yz)
\nn \\
&-& \frac{\Theta(\lambda_p)\sqrt(\lambda_p)}{p^2}
\ln\left|\frac{p^2-y-z+\frac{\lambda_p}{T-p^2}}{p^2-y-z+\sqrt{\lambda_p}}\right|
+\frac{\Theta(\lambda_0)\sqrt{\lambda_0}}{p^2}
\ln\left|\frac{-y-z+\frac{\lambda_0}{T}}{-y-z+\sqrt{\lambda_0}}\right|\,
,
\nn \\
\end{eqnarray}
with the following abbreviations:
\begin{eqnarray}
\lambda_p&=&\lambda(p^2,y,z) ,
\nn \\
\sqrt{\lambda_0}&=&|y-z|\, ,
\nn \\
T&=&(\sqrt{y}+\sqrt{z})^2\, .
\end{eqnarray}
These expressions have been actually used for the numerical evaluation of $M$ in the presented paper.



\end{document}